\documentstyle[prl,aps,floats,epsf]{revtex}

\begin{document} 
\draft

\twocolumn[\hsize\textwidth\columnwidth\hsize\csname
@twocolumnfalse\endcsname

\preprint{HEP/123-qed}

\title{Doping Dependence of the Electronic Structure of 
Ba$_{1-x}$K$_{x}$BiO$_{3}$ Studied by X-Ray Absorption Spectroscopy}

\author{K. Kobayashi, T. Mizokawa, A. Ino, J. Matsuno, A. Fujimori}

\address{Department of Physics, University of Tokyo, Hongo 7-3-1, 
Bunkyo-ku, Tokyo, 113-0033, Japan}

\author{H. Samata, A. Mishiro and Y. Nagata} 

\address{College of Science and Engineering, Aoyama-Gakuin University, 
Chitosedai 6-16-1, Setagaya-ku, Tokyo, 157-8572, Japan}

\author{F. M. F. de Groot}

\address{Heterogeneous Catalysis, Department of Inorganic Chemistry, 
University of Utrecht, 3584 CA Utrecht, the Netherlands}

\date{Submitted to Phys. Rev. B, Feb. 1, 1999}
\maketitle

\begin{abstract}
We have performed x-ray absorption spectroscopy (XAS) and x-ray 
photoemission spectroscopy (XPS) studies of single crystal 
Ba$_{1-x}$K$_{x}$BiO$_{3}$ (BKBO) covering the whole composition range 
$0 \leq x \leq 0.60$.  Several features in the oxygen 1\textit{s} core 
XAS spectra show systematic changes with $x$.  Spectral weight around 
the absorption threshold increases with hole doping and shows a finite 
jump between $x=0.30$ and 0.40, which signals the metal-insulator 
transition.  We have compared the obtained results with band-structure 
calculations.  Comparison with the XAS results of 
BaPb$_{1-x}$Bi$_{x}$O$_{3}$ has revealed quite different doping 
dependences between BKBO and BPBO. We have also observed systematic 
core-level shifts in the XPS spectra as well as in the XAS threshold 
as functions of $x$, which can be attributed to a chemical potential 
shift accompanying the hole doping.  The observed chemical potential 
shift is found to be slower than that predicted by the rigid band 
model based on the band-structure calculations.
\end{abstract}
\pacs{79.60.-i, 71.30.+h,74.25.Jb,74.70.Ad}


\vskip1pc]
\narrowtext

\section{Introduction}
The hole-doped bismuthate Ba$_{1-x}$K$_{x}$BiO$_{3}$ (BKBO) has been 
fascinating researchers since its 
discovery~\cite{MattheissPRB1988,CavaNature1988} for its diverse 
physical properties such as superconductivity, metal-insulator 
transition and charge-density wave (CDW) formation.  The crystal 
structure is a distorted ($0 \leq x < x_{c}$) or cubic ($x_{c} < x$) 
three-dimensional perovskite~\cite{PeiPRB1990}, where $x_{c} \sim 
0.38$.  The parent compound BaBiO$_{3}$ is an insulator in contrast to 
the prediction of band theory that it is a metal with the half-filled 
non-degenerate Bi 6\textit{s}-O 2\textit{p} antibonding band crossing 
the Fermi level 
($E_{F}$)~\cite{MattheissPRB1983,TakegaharaJPSJ1987,MattheissPRL1988,HamadaPRB1989}. 
 This discrepancy between band theory and experiment is accounted for 
as due to a charge disproportionation of Bi into Bi$^{3+}$ and 
Bi$^{5+}$ sites or a CDW formation accompanied by breathing and 
tilting distortions of the BiO$_{6}$ 
octahedra~\cite{MattheissPRB1983,CoxSSC1976,LiechtensteinPRB1991}.  By 
increasing $x$ from 0, i.~e., by hole doping into BaBiO$_{3}$, BKBO 
undergoes a semiconductor-to-superconductor transition at $x = x_{c}$.  
The transition temperature ($T_{C}$) becomes highest $\sim 30$ K at $x 
= x_{c}$ ~\cite{CavaNature1988}, which is the highest among the 
copper-free oxides.  It remains superconducting up to $x \sim 0.6$, 
above which there is a solubility limit of K atoms~\cite{UchidaDoc}.  
Those striking properties have often been compared with those of the 
typical cuprate system La$_{2-x}$Sr$_{x}$CuO$_{4}$ (LSCO), which also 
shows a transition from the insulating La$_{2}$CuO$_{4}$ into the 
superconducting phase with hole doping.  Unlike BKBO, however, the 
crystal structure of LSCO is of the layered perovskite type and the 
parent material La$_{2}$CuO$_{4}$ is an antiferromagnetic insulator 
rather than a CDW insulator.  It is noted that BKBO bears a close 
relationship with the BaPb$_{1-x}$Bi$_{x}$O$_{3}$ (BPBO) system, which 
shows superconductivity between $x = 0$ and $\sim 0.35$ with the 
maximum $T_{C}$ of $\sim 12$ K around $x \sim 
0.25$~\cite{TajimaPRB1987}.  The origin of the superconductivity in 
BKBO as well as in BPBO has not been settled yet: Both systems are 
\textit{sp} electron systems without magnetic ions, excluding the 
possibility of magnetic pairing mechanisms.

Understanding the electronic structure of BKBO is a necessary step to 
elucidate the mechanism of its superconductivity.  So far many studies 
have been performed on BKBO including transport, optical and tunneling 
experiments.  Quite a few photoemission studies have also been 
reported~\cite{RuckmanPRB1989,NagoshiJPCM1992,NamatamePRB1994,ZakharovPRB1997,WagenerPRB1989} 
while there have been relatively few x-ray absorption spectroscopy 
(XAS) studies~\cite{SuguiPRB1991,QvarfordPRB1996,NylenPRB1998}.  XAS 
gives us information about the unoccupied electronic states and, 
therefore, is suitable for studying the effect of hole doping on the 
electronic structure, as has been successfully applied to 
LSCO~\cite{ChenPRL1991}.  Salem-sugui \textit{et 
al.}~\cite{SuguiPRB1991} observed a prepeak in the O 1\textit{s} XAS 
spectra for BKBO with $x = 0.0$, 0.2 and 0.4 and a continuous change 
across the transition $x\sim x_{c}$ was observed.  Qvarford \textit{et 
al.}~\cite{QvarfordPRB1996} also reported result for $x = 0.1$ and 0.4 
and obtained similar results.  A systematic study covering a wider 
concentration range, however, has been lacking.

In this paper, we report on a detailed XAS study of 
Ba$_{1-x}$K$_{x}$BiO$_{3}$ using single crystals covering the entire 
available K concentration range ($0 \leq x \leq 0.60$) for the first 
time.  Gradual and systematic $x$-dependent changes were observed for 
several features in the XAS spectra.  We compare these results with 
the band-structure calculations and with the XAS results of BPBO. A 
quantitative analysis has been made on the spectral weight of the 
prepeak.  Contrasting doping behaviors have been revealed between BKBO 
and BPBO. We also performed XPS measurements on the core levels of 
BKBO. The observed core-level shifts will be discussed in terms of the 
chemical potential shift due to the hole doping.

\section{Experiments}

Ba$_{1-x}$K$_{x}$BiO$_{3}$ single crystals ($x =$ 0.10, 0.20, 0.30, 
0.40, 0.50 and 0.60) were grown by an electrochemical method from a 
KOH solution with Ba$($OH$)_{2}\cdot 8$H$_{2}$O and Bi$_{2}$O$_{3}$.  
The composition of each sample was determined by 
EPMA~\cite{BKBO_preparation}.  A BaBiO$_{3}$ single crystal was 
prepared by the flux method.

XAS measurements were performed at beamline BL-2B of Photon Factory, 
High Energy Accelerator Research Organization.  Photon energies were 
calibrated using the Cu 2$p_{3/2}$ edge of Cu metal at 932.5 
eV~\cite{GrioniPRB1989} and the O 1\textit{s} edge of LaCoO$_{3}$ at 
529.3 eV~\cite{SaitohPRB1997}.  O 1\textit{s} XAS spectra were taken 
by the total electron-yield method with photoelectrons.  The energy 
resolution including the lifetime of the O 1\textit{s} core hole was 
$\sim 0.7$ eV at 530 eV. The samples were cooled down to $\sim 50$ K 
using a closed-cycle He refrigerator.  The pressure in the 
spectrometer was $\sim 4 \times 10^{-10}$ Torr.

XPS measurements were performed using the Mg K$\alpha$ line 
($h\nu=1253.6$ eV) and photoelectrons were collected using a PHI DCMA. 
Calibration and estimation of the instrumental resolution were done 
using gold evaporated on the sample surface defining Au $4f_{7/2} = 
84.0$ eV. The total resolution was $\sim 1$ eV, including both the 
light source and the instrumental resolution.  The base pressure in 
the spectrometer was $\sim 6 \times 10^{-10}$ Torr.  All the XPS 
measurements were performed at room temperature in order to avoid 
charging effect in the insulating phase.

The samples were scraped \textit{in situ} with a diamond file for both 
measurements.  For the XAS measurements, which are less 
surface-sensitive than XPS, samples were scraped until the spectra 
showed no further change.  The obtained XAS spectra seem consistent 
with the previous reports for $x=0$~\cite{deGrootPRB1991} and $x = 
0.40$~\cite{NylenPRB1998}.  For the XPS measurements, scraping was 
performed until the O 1\textit{s} core-level spectra became a single 
peak.  We confirmed that the measured peak intensity of the Ba 
3\textit{d} XPS core level was proportional to $x$, while those of the 
O 1\textit{s} and Bi 4\textit{f} core levels were almost independent 
of $x$.

\section{Results and Discussion}

\subsection{Overall Electronic Structure of the Unoccupied States}

\begin{figure} [ht] \center \epsfxsize=85mm \epsfbox{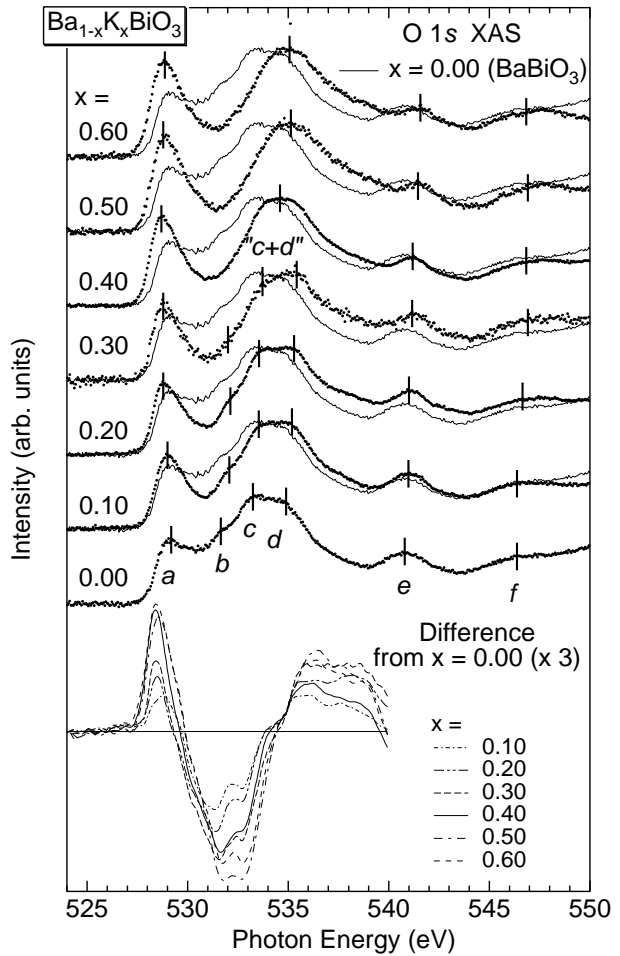} 
\vspace{2pt} \caption{Oxygen 1\textit{s} x-ray-absorption spectra 
(XAS) of a series of Ba$_{1-x}$K$_{x}$BiO$_{3}$ ($0 \leq x \leq 
0.60$).  The spectrum of BaBiO$_{3}$ is superimposed on each spectrum 
as a solid curve.  They are normalized to their peak height around 
$h\nu = 533$--535 eV. The positions of \textit{a}--\textit{e} are 
plotted in Fig.~\ref{BKBO_XAS_PeakPosition}.  In the bottom panel are 
shown the difference spectra from the XAS spectrum of BaBiO$_{3}$.}
\label{BKBO_XAS_WideScan}
\end{figure}

\begin{figure} [ht] \center \epsfxsize=70mm \epsfbox{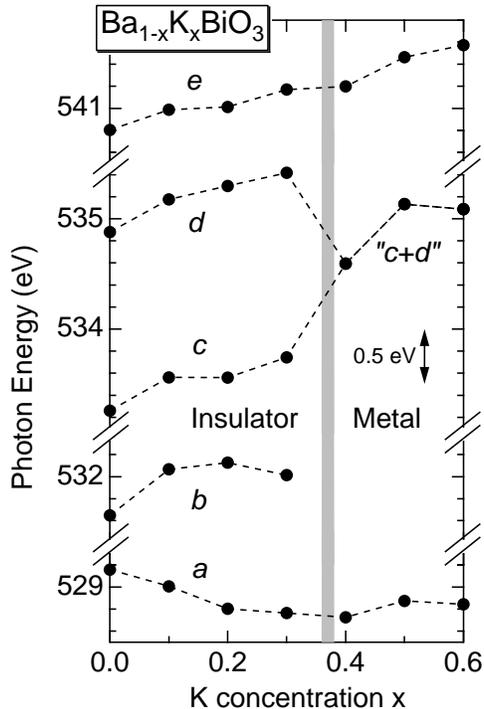} 
\vspace{2pt} \caption{Positions of structures \textit{a}--\textit{e} 
indicated in Fig.~\ref{BKBO_XAS_WideScan}.}
\label{BKBO_XAS_PeakPosition} 
\end{figure}

Figure~\ref{BKBO_XAS_WideScan} shows the XAS spectra of 
Ba$_{1-x}$K$_{x}$BiO$_{3}$ for various $x$'s.  We superimpose the 
spectrum of BaBiO$_{3}$ ($x =0.00$) on each spectrum ($x \geq 0.10$) as a 
solid curve.  Normalization of all the spectra was performed by peak 
height at the photon energy $h\nu = 533$--535 eV. In the bottom, 
difference spectra from the BaBiO$_{3}$ spectrum are shown.  Notable 
structures found in the spectra are labeled as \textit{a}--\textit{f} 
in Fig.~\ref{BKBO_XAS_WideScan}, and the positions of structures 
\textit{a}--\textit{e} are plotted in 
Fig.~\ref{BKBO_XAS_PeakPosition}.

The O 1\textit{s} XAS spectra represent the unoccupied partial density 
of states (PDOS) of \textit{p} character at the oxygen site.  
Structure \textit{a} near the absorption edge $h\nu \sim 528.0$--528.5 
eV grows almost monotonously as $x$ increases, which is also 
substantiated in the bottom panel of Fig.~\ref{BKBO_XAS_WideScan}.  At 
$x = 0.00$, the shape of the prepeak \textit{a} is rounded, while it 
becomes a sharper peak at $x \geq 0.10$.  For the spectra $0.10 \leq x 
\leq 0.30$, where BKBO is insulating, the peak grows slowly as $x$ 
increases.  At $x = 0.40 > x_{c}$, where BKBO enters the metallic 
region, the prepeak suddenly becomes intense and keeps its height 
until $x = 0.60$.  The increase of the prepeak is due to the doped 
holes and the rapid growth from $x=0.30$ to 0.40 would be attributed 
to a transfer of spectral weight to the region near the chemical 
potential, signaling the metal-insulator transition.  This behavior is 
essentially different from that in LSCO~\cite{ChenPRL1991}, where new 
spectral weight appears well below the first peak of the insulator 
La$_{2}$CuO$_{4}$.  A quantitative discussion about the spectral 
weight of the prepeak will be made in Sec.~\ref{PrepeakArea}.

Next, we investigate the behaviors of structures 
\textit{b}--\textit{f}.  Structure \textit{b}, which appears as a 
shoulder-like structure for $x = 0.00$ at $h\nu \sim 532$ eV, becomes 
obscure in going from $x = 0.00$ to 0.30 and disappears at $x \geq 
0.40$.  Hence, \textit{b} may be attributed to Ba character, which 
point will be discussed below.  The behavior of structure \textit{c} 
at $h\nu \sim 533$--534 eV and structure \textit{d} at $h\nu \sim 535$ 
eV is also systematically dependent on $x$.  At $x = 0.00$, feature 
\textit{c} with a slightly smaller feature \textit{d} is observed.  
This two-component peak centered around $h\nu \sim 534$ eV is 
monotonously shifted to higher photon energies with $x$, until the 
center of the peak reaches $h\nu \sim 535$ eV. At the same time, 
\textit{d} becomes stronger than \textit{c} at $x \sim 0.30$ and, 
finally, structures \textit{c} and \textit{d} merge at $x \geq 0.40$ 
into a peak as labelled ``$c+d$'' in Figs.~\ref{BKBO_XAS_WideScan} and 
\ref{BKBO_XAS_PeakPosition}.  This observation can be explained as 
follows: Structures \textit{c} and \textit{d} are mainly of Ba 
character at $x=0.00$, while the substituted states of K character 
appear at the position of \textit{d} for $x \geq 0.10$ and, therefore, 
the gradual change results from the change of the electronic structure 
due to the substitution of K for Ba.  As will be discussed in 
Sec.~\ref{SUB_BPBO_XAS}, this is also supported by the result of the 
XAS spectra of BPBO~\cite{deGrootPRB1991}, where the corresponding 
\textit{c} structure does not move to higher photon energies so much 
as that in BKBO.

Besides \textit{a}--\textit{d}, the behaviors of the structures in the 
higher photon energy region are worth mentioning although they behave 
less drastically than those near the absorption edge.  On one hand, 
structure \textit{e} around $h\nu \sim 541$ eV simply becomes weak 
with $x$, indicating that this structure is of Ba character.  On the 
other hand, the broader structure \textit{f} around $h\nu \sim 547$ eV 
does not change in the entire $x$ range, indicating this to be of Bi 
or purely of O character.

It is interesting to compare these observations with the previous 
report on the unoccupied states of BKBO studied by inverse 
photoemission spectroscopy.  Wagener \textit{et 
al.}~\cite{WagenerPRB1989} assigned structures at about 4, 7, 9 and 14 
eV above $E_{F}$ (corresponding structures \textit{b}--\textit{e}) to 
Bi 6\textit{p}, Ba 5\textit{d}, K 3\textit{d} and Ba 4\textit{f} 
states, respectively.  As for structures \textit{c}, \textit{d} and 
\textit{e}, their assignment agrees with ours.  If \textit{b} is 
assigned to the Bi 6\textit{p} band as they identified, our 
observation that \textit{b} disappears at $x \geq 0.40$ may indicate a 
change of the electronic structure caused by the change of the crystal 
structure.

\begin{figure} [ht] \center \epsfxsize=75mm \epsfbox{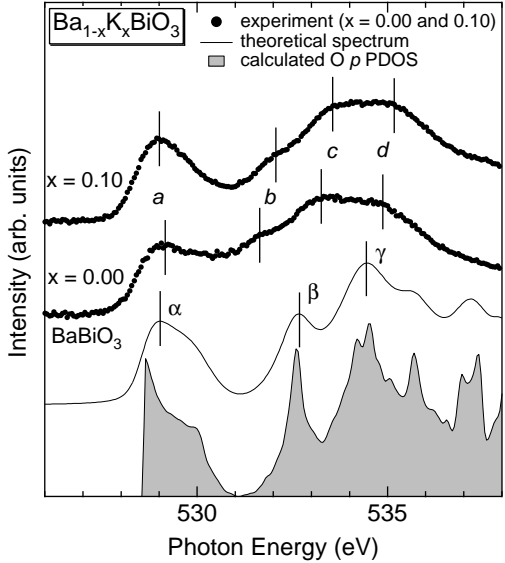} 
\vspace{2pt} \caption{Comparison of the XAS spectra of BaBiO$_{3}$ and 
Ba$_{0.9}$K$_{0.1}$BiO$_{3}$(dots) with the theoretical spectrum 
(solid curves) derived from the unoccupied O \textit{p} 
PDOS~\protect\cite{TakegaharaJPSJ1987}.  The labels 
\textit{a}--\textit{d} are the same as in 
Fig.~\ref{BKBO_XAS_WideScan}.}
\label{Exp_vs_Band}
\end{figure}

In Fig.~\ref{Exp_vs_Band}, we show comparison between the XAS spectra 
of BaBiO$_{3}$ and Ba$_{0.9}$K$_{0.1}$BiO$_{3}$ and the theoretical 
spectrum of BaBiO$_{3}$ derived from the unoccupied oxygen \textit{p} 
PDOS~\cite{TakegaharaJPSJ1987}.  The oxygen PDOS has been broadened 
with a Gaussian and a Lorentzian which represent the instrumental 
resolution $\sim 0.5$ eV and the lifetime broadening of O 1\textit{s} 
core hole $\sim 0.3$ eV, respectively~\cite{deGrootPRB1991}.  The 
energy scale of the theoretical spectrum has been shifted so as to 
compare well with the experimental spectra.  Structures in the 
theoretical spectrum are labelled as $\alpha$--$\gamma$.  Although the 
lineshapes of the XAS spectra tend to be distorted by the core-hole 
potential, the theoretical spectrum reproduces experiment to some 
extent: Structures $\alpha$, $\beta$ and $\gamma$ seem to correspond 
to \textit{a}, \textit{b} and \textit{c} (or \textit{d}), 
respectively.  In addition, it is interesting to note that, as for the 
prepeak \textit{a}, the theoretical spectrum resembles the 
experimental spectrum of Ba$_{0.9}$K$_{0.1}$BiO$_{3}$ rather than that 
of BaBiO$_{3}$.  Because both samples are insulating, this cannot be 
explained only by the effect of the core-hole potential.  This may be 
related to the difference of the crystal structure between $x = 0.00$ 
and $x =0.10$ since the strong breathing-type distortion and the 
tilting of the BiO$_{6}$ octahedra exist for $x \lesssim 0.1$ with a 
large electron-phonon coupling 
constant~\cite{PeiPRB1990,LiechtensteinPRB1991}.

\subsection{Comparison with the XAS Spectra of 
BaPb$_{1-x}$Bi$_{x}$O$_{3}$}
\label{SUB_BPBO_XAS}
\begin{figure} [ht] \center \epsfxsize=85mm \epsfbox{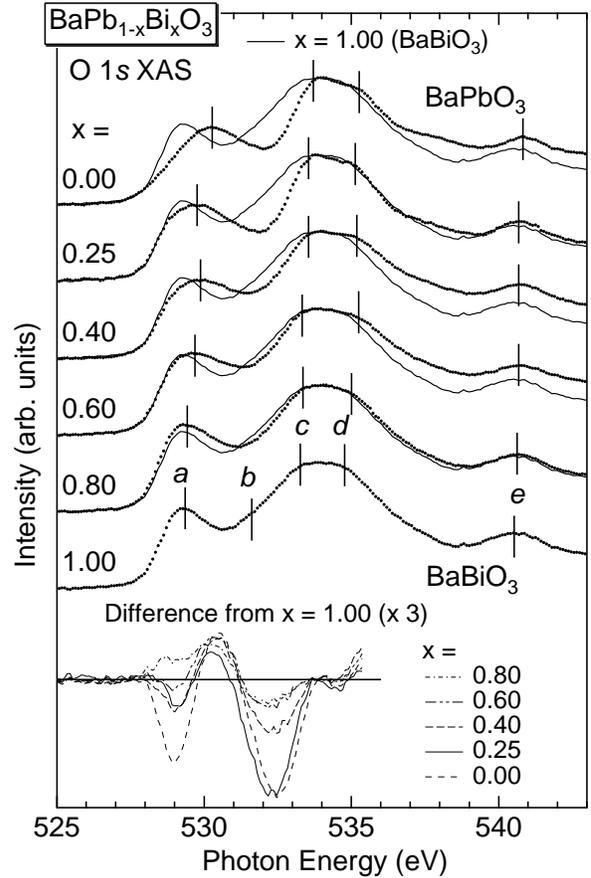} 
\vspace{2pt} \caption{O 1\textit{s} XAS spectra of a series of 
BaPb$_{1-x}$Bi$_{x}$O$_{3}$ ($0.0 \leq x \leq 1.0$).  The spectrum of 
BaBiO$_{3}$ is superimposed on each spectrum.  They are normalized to 
their peak height around $h\nu = 534$ eV. The positions of 
\textit{a}--\textit{e} are plotted in 
Fig.~\ref{BPBO_XAS_Peakposition}.  In the bottom panel are shown the 
difference spectra from that of BaBiO$_{3}$.}
\label{BPBO_XAS}
\end{figure}

\begin{figure} [ht] \center \epsfxsize=70mm 
\epsfbox{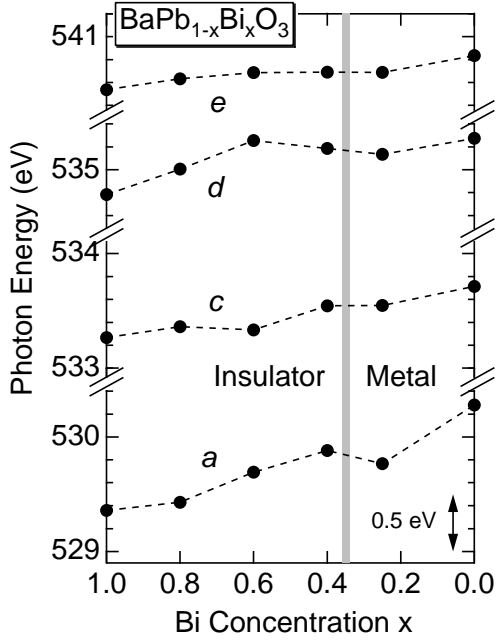} \vspace{2pt} \caption{Positions of 
structures \textit{a} and \textit{c}--\textit{e} indicated in 
Fig.~\ref{BPBO_XAS}.}
\label{BPBO_XAS_Peakposition}
\end{figure}

Next we compare our results with the previous report on the XAS spectra
of the BPBO system~\cite{deGrootPRB1991}.  In the upper panel of 
Fig.~\ref{BPBO_XAS}, the XAS spectra of a series of 
BaPb$_{1-x}$Bi$_{x}$O$_{3}$ ($0.0 \leq x \leq 1.0$) are shown by dots, 
with the BaBiO$_{3}$ XAS spectrum superimposed for 
comparison~\cite{CommentOnBPBO}.  All the spectra have been normalized 
to their peak height around $h\nu \sim 534$ eV. We have labelled 
structures as \textit{a}--\textit{e}.  In the bottom, the difference 
spectra from the BaBiO$_{3}$ spectrum are shown.  The positions of 
structures \textit{a}--\textit{e} are plotted in 
Fig.~\ref{BPBO_XAS_Peakposition}.

Although structure \textit{b} is only weekly visible, structures 
\textit{c}, \textit{d} and \textit{e} correspond well to those of BKBO 
in Fig.~\ref{BKBO_XAS_WideScan}.  Unlike BKBO, \textit{c} and 
\textit{d} do not greatly change their positions and lineshapes with 
$x$.  \textit{e} also does not change its position with $x$.  This 
supports the above assignment for BKBO that \textit{c}, \textit{d} and 
\textit{e} are mainly of Ba character.  On the other hand, the 
position and lineshape of the prepeak labelled \textit{a} changes 
dramatically with $x$, while their intensity relative to the main peak 
at $h\nu = 533$--535 eV is conserved.  This observation is clearly 
contrasted with the case of BKBO, where the intensity of the prepeak 
increases with $x$.  The difference spectra in the bottom panel also 
clarify this point.  As for BPBO, the intensity of the prepeak 
\textit{a} a little increases in going from $x = 1.00$ to 0.80, and 
then decreases almost monotonously as $x$ decreases.  Unlike BKBO, 
there is another structure in the difference spectra at $h\nu \sim 
530$ eV, which does not behave systematically, while the broad dip at 
$\sim 532$--533 eV increases with decreasing $x$.  Such complicated 
behaviors significantly differ from those of BKBO, reflecting the 
different doping processes between BPBO and BKBO. In the former 
system, the Bi 6\textit{s}-O 2\textit{p} antibonding band is replaced 
by the Pb 6\textit{s}-O 2\textit{p} antibonding band near $E_{F}$.  In 
the latter system, the doped holes are rather simply accommodated in 
the Bi 6\textit{s}-O 2\textit{p} antibonding band.  The different 
doping processes are also reflected on the opposite shifts of peak 
\textit{a} in BKBO and BPBO (Figs.~\ref{BKBO_XAS_PeakPosition} and 
\ref{BPBO_XAS_Peakposition}).  Other structures are shifted more 
slowly in BPBO than in BKBO.

\subsection{Electronic Structure near the Fermi Level}
\label{PrepeakArea}

\begin{figure} [ht] \center \epsfxsize=80mm \epsfbox{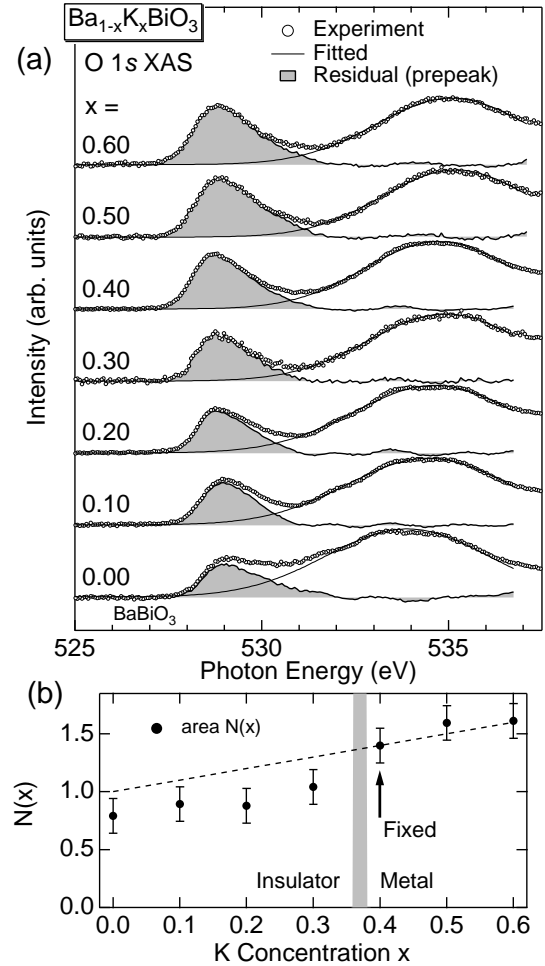} 
\vspace{2pt} \caption{(a) Prepeak of the XAS spectra extracted on the 
assumption that the peak at $h\nu = 534$--535 eV is a Gaussian.  (b) 
Area $N(x)$ of the prepeak relative to that of the main peak.  $N(x)$ 
has been normalized assuming that $N(x=0.40) = 1.40$.  The dashed line 
represents the ``ideal hole doping'' as explained in the text.}
\label{BKBO_GaussFit}
\end{figure}

In this section, we discuss how doped holes are accommodated in 
the Bi 6\textit{s}-O 2\textit{p} antibonding band in BKBO based on the 
doping dependence of the intensity of the prepeak.  In order to 
characterize how the prepeak grows with K substitution, the peak 
around $h\nu = 534$--535 eV has been subtracted by assuming it to be a 
Gaussian, as shown in Fig.~\ref{BKBO_GaussFit} (a).  Here, the 
shaded area represents the extracted prepeak.  In 
Fig.~\ref{BKBO_GaussFit} (b), the area $N(x)$ of the prepeak relative 
to that of the Gaussian is plotted as a function of $x$.  We have 
scaled $N(x)$ so that $N(x=0.40)=1.40$ because $N(x)$ would be 
proportional to the number of empty states in the Bi 6\textit{s}-O 
2\textit{p} antibonding band if the strength of the Bi 6\textit{s}-O 
2\textit{p} hybridization does not change with $x$.

In accordance with the observation in Fig.~\ref{BKBO_XAS_WideScan} 
that the prepeak grows systematically with $x$, the extracted $N(x)$ 
is a monotonously increasing function of $x$ in the whole $x$ range.  
The dashed line in Fig.~\ref{BKBO_GaussFit} (b) represents the ``ideal 
hole doping'' $N(x) = 1 + x$.  One can see a deviation from this 
``ideal'' behavior below $x=0.30$ with a jump between $x=0.30$ and 
$0.40$, i.~e., below and above $x = x_{c}$.  The present result 
suggests that below $x = x_{c}$, doped holes may not be accommodated 
in the Bi 6\textit{s}-O 2\textit{p} antibonding band in the same way 
as in the metallic phase, resulting in the reduced $N(x)$ and the 
insulating behavior.  In other words, spectral weight due to the empty 
Bi 6\textit{s}-O 2\textit{p} states, which should be proportional to 
$1+x$, is not concentrated near $E_{F}$, but is also distributed away 
from $E_{F}$, presumably overlaid by the Gaussian.

\begin{figure} [ht] \center \epsfxsize=85mm \epsfbox{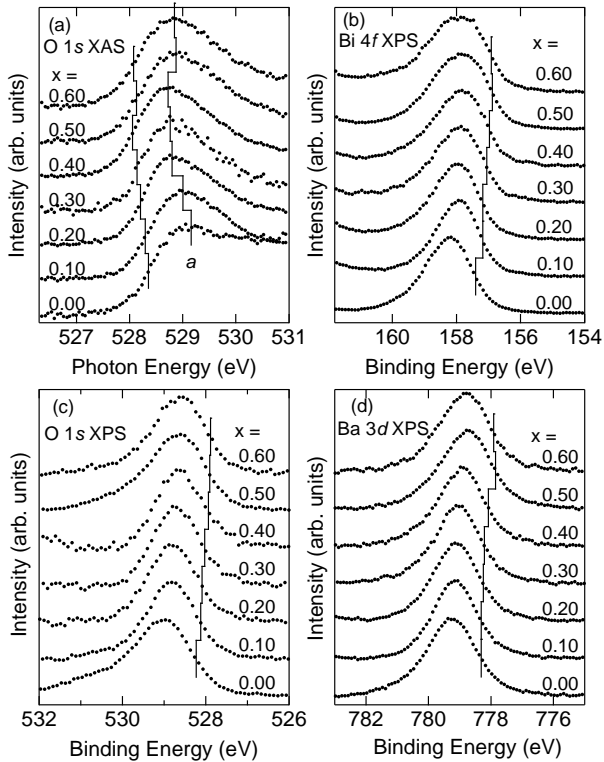} 
\vspace{2pt} \caption{(a) O 1\textit{s} XAS spectra of BKBO near the 
absorption threshold.  The threshold defined as the midpoint of the 
peak are marked with vertical bars.  The position of the prepeak 
\textit{a} is also shown.  (b) (c) and (d) are the XPS spectra of the 
Bi 4\textit{f}, O 1\textit{s} and Ba 3\textit{d} core levels.  The 
midpoint of the peak is marked with the vertical bars.  In each panel, 
the spectra have been normalized to the peak height.}
\label{XAS_XPS_Shift}
\end{figure}

Next, we focus on the shift of the threshold of the XAS spectra.  
Figure~\ref{XAS_XPS_Shift} (a) shows the XAS spectra of BKBO near the 
absorption threshold.  We define the threshold by the midpoint of the 
leading edge as marked by the vertical bars in the figure.  The result 
of the threshold shift is shown in the bottom panel of 
Fig.~\ref{Result_Shift}.  The positions of the prepeak \textit{a} 
around $h\nu \sim 529$ eV are also indicated by vertical bars in 
Fig.~\ref{XAS_XPS_Shift} (a).  In going from $x = 0.00$ to 0.20, the 
prepeak position rapidly moves toward lower photon energies as shown 
in Fig.~\ref{BKBO_XAS_PeakPosition}.  On the other hand, the threshold 
is shifted more slowly than that of the prepeak in going from $x = 
0.00$ to $x = 0.20$, meaning that the prepeak becomes broader with 
decreasing $x$.  This may have an origin common with the depression of 
the spectral weight $N(x)$ in the insulating phase and may be 
attributed to the structural distortion and the CDW gap formation in 
the insulating phase.  It has been reported in the optical study that 
the CDW gap of $\sim 2$ eV opens at $x = 0.00$, while it almost 
collapses already at $x = 0.24$~\cite{KarlowPRB1993}.  On the other 
hand, in the metallic region at $x \geq 0.40$, the shift of the XAS 
threshold seems almost saturated.  This saturation may arise from the 
difficulty in the definition of the threshold in the metallic region 
because of the pile-up of spectral weight near $E_{F}$, since the O 
1\textit{s} XPS peak shift behaves systematically for $x \leq 0.50$ as 
reported below.

The depression of $N(x)$ in the insulating phase implies a possible 
failure of the simple rigid-band model.  In this region, doped holes 
would be localized due to strong electron-phonon coupling, which 
decreases the spectral weight near $E_{F}$.  This is consistent with 
the report by Puchkov \textit{et al.}~\cite{PuchkovPRB1996} that 
spectral weight observed in the mid-infrared absorption of the 
insulating BKBO is suppressed and that the lost spectral weight is 
transferred to a higher energy region.  In contrast, above $x = 
x_{c}$, spectral weight of the holes introduced by the substitution of 
K for Ba are distributed near $E_{F}$, leading to the regular behavior 
of $N(x)$.

\subsection{Core-level XPS Spectra and Chemical Potential Shift}
\begin{figure} [ht] \center \epsfxsize=85mm \epsfbox{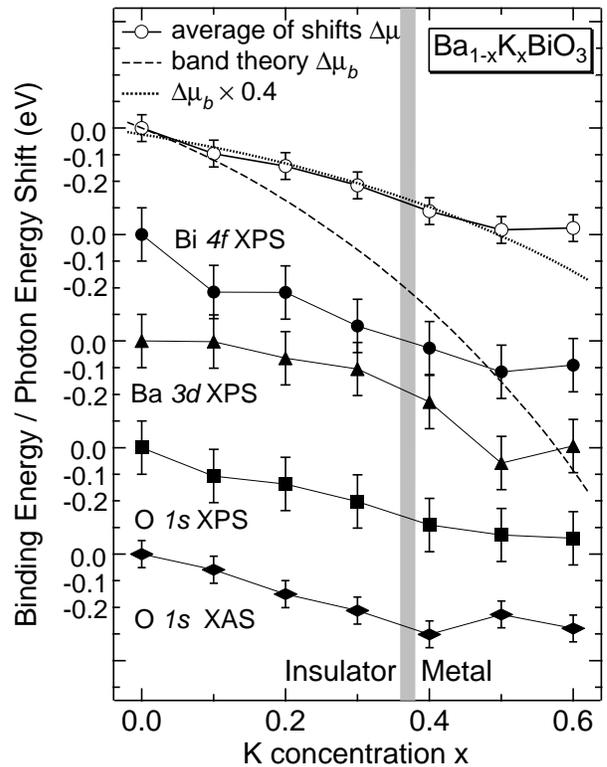} 
\vspace{2pt} \caption{Shifts of the O 1\textit{s} XAS threshold and 
the O 1\textit{s}, Ba 3\textit{d} and Bi 4\textit{f$_{7/2}$} XPS core 
levels.  In the top, the average shift of the XAS threshold and the 
XPS core levels and the theoretical chemical potential shift 
calculated by band theory~\protect\cite{TakegaharaJPSJ1987} assuming 
the rigid band model are shown.}
\label{Result_Shift}
\end{figure}

In Figs.~\ref{XAS_XPS_Shift} (b), (c) and (d), we also show the 
results of the Bi 4\textit{f$_{7/2}$}, O 1\textit{s} and Ba 
3\textit{d} core-level XPS spectra, respectively.  According to the 
high resolution XPS study~\cite{QvarfordPRB1996}, the lineshapes of 
the core levels of the metallic BKBO are not single peaks but of 
multi-components with an energy-loss structure on the high binding 
energy ($E_{B}$) side.  In fact, the core-level spectra in the 
metallic phase (particularly, $x = 0.50$ and 0.60) show somewhat 
broader peak widths than those in the insulating phase.  In order not 
to be disturbed by such energy-loss structures, we use the midpoint of 
the lower binding energy edge of each peak.  The peak positions thus 
obtained are shown by vertical bars in Figs.~\ref{XAS_XPS_Shift} (b), 
(c) and (d).  The results of the Bi 4\textit{f$_{7/2}$}, O 1\textit{s} 
and Ba 3\textit{d} core-level binding energies relative to $x = 0.00$ 
are compiled in Fig.~\ref{Result_Shift}.  In going from $x = 0.00$ to 
0.50~\cite{CommentOn060}, each core level gradually moves to lower 
$E_{B}$ by $\sim 0.3$--0.5 eV. The direction and amount of the 
core-level shifts are common to all the core levels in agreement with 
the previous result~\cite{NamatamePRB1994}.  Between $x = 0.00$ and 
0.40, the shift of the O 1\textit{s} core-level XPS spectrum and the 
shift of the threshold of the O 1\textit{s} XAS spectrum agree with 
each other quite well, indicating that the XAS threshold corresponds 
to a transition from the O 1\textit{s} core level to states at 
$E_{F}$.

Here, it is important to note that the core levels of both anion (O) 
and cations (Ba, Bi) move in the same direction, indicating that the 
observed shift is a measure of the chemical potential shift caused by 
the hole doping.  We have obtained the average of the shifts of the O 
1\textit{s} XAS threshold and the XPS core levels of Ba 3\textit{d}, O 
1\textit{s} and Bi 4\textit{f} as plotted by open circles in 
Fig.~\ref{Result_Shift} and regard it as the experimentally deduced 
chemical potential shift $\Delta\mu(x)$.  Except for $x=0.60$, 
$\Delta\mu(x)$ decreases monotonously.  In Fig.~\ref{Result_Shift}, we 
have also plotted the chemical potential shift $\Delta\mu_{b}(x)$ 
predicted by the band-structure calculations based on the rigid band 
model~\cite{TakegaharaJPSJ1987} by a dashed curve.  The qualitative 
behavior of $\Delta\mu(x)$ implies that the doping dependence of the 
BKBO system obeys the rigid band model, as pointed out before by 
Namatame \textit{et al.}~\cite{NamatamePRB1994}.  Quantitatively, 
however, the amount of $\Delta\mu(x)$ is reduced compared to 
$\Delta\mu_{b}(x)$, indicating that the rigid band model based on the 
band-structure calculations does not succeed in the quantitative 
explanation of $\Delta\mu(x)$.  In fact, we found that $0.4 \times 
\Delta\mu_{b}(x)$ well reproduces the $\Delta\mu(x)$ as shown by a 
dotted line in the figure.  Zakharov \textit{et 
al.}~\cite{ZakharovPRB1997} observed a 0.35 eV shift of the O 
2\textit{p} band between $x =0.1$ and 0.4 by photoemission 
spectroscopy, corresponding to the factor of $\sim 0.6$.  Namatame 
\textit{et al.}~\cite{NamatamePRB1994} also reported that the measured 
chemical potential shift is reduced almost to half of the theoretical 
prediction.

Finally, it should be noted that the behavior of $\Delta\mu(x)$ is 
gradual without any abruptness across $x=x_{c}$, which is in clear 
contrast to the fact that there is a finite jump in the prepeak area 
$N(x)$ between $x=0.30$ and 0.40.  These two facts may be related to 
the opening of the pseudogap in the metallic 
samples~\cite{NamatamePRB1994}.  This is an issue to be clarified in 
future.

\section{Conclusion}
We have performed XAS and XPS measurements of BKBO in the whole $x$ 
range ($0 \leq x \leq 0.60$).  The XAS results show that the overall 
electronic structure of the unoccupied states changes systematically 
with substitution of K for Ba.  That is, the intensity of the prepeak 
increases monotonously with a finite jump between $x = 0.30$ and 0.40, 
and the absorption threshold moves toward lower energies, at least in 
the insulating phase.  Good overall agreement with the band-structure 
calculations has been obtained.  Comparison with the XAS result of the 
BPBO system has revealed very different doping-dependent changes in 
the electronic structure between BKBO and BPBO. In the latter system, 
Pb substitution replaces the prepeak in BaBiO$_{3}$ of Bi 
6\textit{s}-O 2\textit{p} anti-bonding character by another peak of Pb 
6\textit{s} character at a higher energy.  The XPS measurements reveal 
that the chemical potential shift $\Delta\mu(x)$ derived from the 
core-level shifts moves in the direction predicted by the rigid-band 
model.  The amount of the shift, however, is almost half of that 
predicted by the band-structure calculations.

\section{Acknowledgements}
The authors would like to thank Y. Azuma, T. Miyashita and the staff 
of Photon Factory for technical support in the XAS measurements.  We 
would like to thank H. Takagi for providing us with the BaBiO$_{3}$ 
sample and T. Hashimoto for informative discussions.  This work was 
supported by a Special Coordination Fund from the Science and 
Technology Agency of Japan.  One of us (KK) is supported by a Research 
Fellowship of the Japan Society for the Promotion of Science for Young 
Scientists.  The experiment at Photon Factory was done under the 
approval of the Photon Factory Program Advisory Committee (Proposal 
No.  94-G361).


\begin{references}
\bibitem{CommentOn060}
The singular behavior at $x=0.60$ might be due to the fact that the sample is 
near the solubility limit of K in the BKBO system.
\end{references}
\end{document}